\documentstyle[12pt]{article}

\newcommand{\be}{\begin{equation}}
\newcommand{\ee}{\end{equation}}
\newcommand{\beqa}{\begin{eqnarray}}
\newcommand{\eeqa}{\end{eqnarray}}

\begin{document}

\title{Complete Model of a Self-gravitating Cosmic String I.
	   A New Class of Exact Solutions and Gravitational Lensing}
\author{
Charles C.~Dyer \\
{\it Scarborough College and Department of Astronomy, University
of Toronto,} \\
{\it 1265 Military Trail, Scarborough, Ontario, Canada M1C 1A4} \\
and \\
Francine R. Marleau \\
{\it Department of Astronomy, University of Toronto,} \\
{\it 60 St. George Street, Toronto, Ontario, Canada M5S 1A7}
}
\date{}
\maketitle

\begin{abstract}
We find solutions of Einstein's field equation for topologically stable
strings associated with the breaking of a U(1) symmetry.  Strings form in
many GUTs and are expected whenever the homotopy group $\Pi_1(M_0)$ is
non-trivial.  The behavior of the fields making up the string is
described by the Euler-Lagrange equations.  These fields appear in the
energy-momentum tensor so we must solve simultaneously for the coupled
Einstein-scalar-gauge field equations.  Numerical results are obtained
using a Taylor-series method.  We obtain a 5-parameter family of solutions
and discuss their physical characteristics.  Significant gravitational
lensing can occur due to strings based on this model and are shown for
different solutions.  Finally, we prove that the assumption of regularity
at the string axis is not necessary by looking at the physical properties
of the string solutions.
\end{abstract}

\newpage
\section{Introduction}
Phase transitions occur in the early universe as a
consequence of its expansion and cooling.  The transitions cannot be
observed directly but can be inferred from the theory of groups and
symmetries in elementary particle physics.  If these phase transitions
do occur, there will necessarily form topological
defects during the transition.  Theses defects take their name after their
characteristic of being trapped regions of ``old symmetry'' surrounded by
``new symmetry''.  In essence, they retain the characteristics of the state
of the universe as it was before the phase transition.  The topology of the
defect varies according to the symmetry group G characterizing the
fields present in the universe before the symmetry breaking and the
symmetry group H which describes the symmetry of the field after the
symmetry breaking, i.e. H includes all elements of G which leave the vacuum
expectation value of the scalar field invariant.
When the first homotopy group is nontrivial,
$\Pi_1(G/H)\neq I$, i.e. the topological knot cannot be ``unwound'', the
topology of the defect is linear, or string like.  Topologically stable
strings occur in non-Abelian gauge theories such as SU(5) (Shafi and
Vilenkin \cite{shafi}) and SO(10) (Kibble {\it et al.} \cite{lazarides})
grand unified theories but also in the simple case of Abelian U(1) symmetry.

The gravitational field of strings has been studied in general relativity
starting with the work of Vilenkin \cite{vilenkin}, subsequent to the
Newtonian approach to topological defects given by
Zel'dovich {\it et al.} \cite{zeldovich} and Kibble \cite{kibble}.
He used the linear
approximation of general relativity and an energy-momentum tensor
which has no lateral stresses but only terms describing the energy density
and the pressure (tension) along the axis of the string.  This assumption
was used later in work such as that of Gott \cite{gott}
and Hiscock \cite{hiscock} in looking
for exact solutions of Einstein's equations for a string where, in
addition, the energy
density of the string was taken to be constant.  It has since been shown
by Raychaudhuri \cite{raychaudhuri} that the Gott and Hiscock
solution is not consistent with proper boundary conditions.

Subsequently, there have been a number of attempts to obtain better models
of cosmic strings.  These range from treatments which impose a fixed
background geometry wherein the properties of the string are calculated,
to the attempt of Laguna-Castillo and Matzner \cite{matzner} and
Garfinkle and Laguna \cite{laguna} who alternately
held the metric functions fixed while integrating the string equations
and held the string field properties fixed while computing the new metric
from the Einstein equations.

The purpose of this paper is to extend previous attempts at describing
gravitating cosmic strings by simultaneously solving the coupled
Einstein-scalar-gauge field equations, so that our model will include
the effect of the energy-momentum of the string on the background.
In addition, we want to study the physical properties of the string
solutions that emerge from this set of differential equations.
Recently, Shaver \cite{shaver} has examined the equations for non-stationary
cosmic strings and solved them for the simple energy-momentum tensor
introduced by Vilenkin.  Based on the general set of equations obtained
by Shaver and dropping the Vilenkin restrictions to the form of the
energy-momentum tensor, we will use a numerical integration technique
to attempt to find all possible solutions for self-gravitating strings.

\section{The coupled Einstein-scalar-gauge field equations}
We will be studying string topological defects associated with the
spontaneous symmetry breaking of an Abelian group G=U(1).
The Lagrangian of this Abelian-Higgs model is:
\be
L = -\frac{1}{2}(D^{\mu}\Phi)^{\ast}(D_{\mu}\Phi) - V(\Phi) -
\frac{1}{2}F_{\mu\nu}F^{\mu\nu} \label{lag}
\ee
where $F_{\mu\nu} = \nabla_{\mu}A_{\nu} - \nabla_{\nu}A_{\mu}$,
$D_{\mu} = \nabla_{\mu} + ieA_{\mu}$, $V(\Phi)$ is the potential of
the scalar field, $A_{\mu}$ is the gauge field, $e$ the gauge coupling
constant, and $\Phi$ the scalar field.  The symmetry breaking
potential has the form $V(\Phi) = \lambda(\Phi^{\ast}\Phi - \eta^{2})^{2}$
where $\lambda$ is the self coupling
constant of the Higgs field and $\eta$ is the value of the
symmetry breaking Higgs field.  Shaver \cite{shaver} has shown that for
the specific choice of the energy-momentum tensor given by Vilenkin
\cite{vilenkin}, $\lambda = e^{2}/8$.  Since we are
concerned with finding solutions where
$T^{r}_{r}$ and $T^{\phi}_{\phi}$ are not necessarily zero, we will set
$\lambda = \alpha e^{2}/8$ where $\alpha$ is a constant and can be
taken as a free parameter.  Note that, when acting on a scalar field,
$\nabla_{\mu}$ simply becomes $\partial_{\mu}$.  Also, due to the
symmetries of the Christoffel symbols, $\nabla_{\mu}A_{\nu} -
\nabla_{\nu}A_{\mu} = \partial_{\mu}A_{\nu} - \partial_{\nu}A_{\mu}$.

The spontaneous symmetry breaking is obtained by introducing the Higgs
field.  The symmetry of the system after such a breaking is then
determined by the degeneracy of the vacuum expectation value of the
scalar field:
\beqa
\Phi & = & |\Phi|e^{i\theta}\;,\;\;0\leq\theta\leq2\pi \\
& = & Re^{i\theta} \label{scalar}
\eeqa
where $R \equiv |\Phi|$.

To describe the gauge field, we refer to the work of
Nielsen and Olesen \cite{nielsen} who obtain from their
second equation of motion for flat space-time:
\beqa
A_{\mu} & = & \frac{1}{e^2}\frac{j_{\mu}}{|\Phi|^2} -
\frac{1}{e}\partial_{\mu}\theta \\
& = & \frac{1}{e}[P - 1] \phi_{,\mu} \label{gauge}
\eeqa
where $j_{\mu} \equiv \partial^{\nu}F_{\mu \nu}$,
$P \equiv j_{\mu} / e|\Phi|^2$, and we have set
$\theta = \phi$, the angular coordinate (see below).  (We can now
rewrite the equation for the scalar field as $\Phi = Re^{i\phi}$.)
These expressions for $A_{\mu}$ and $\Phi$ will be used in our equations
below and we will use $R$ and $P$ to characterize the scalar and gauge
field respectively.

The equations of motion are given by:
\be
\frac{\partial L}{\partial \chi} = \nabla_{\mu}\frac{\partial L}{\partial
(\partial_{\mu}\chi)}
\ee
where $\chi$ is replaced by $R$, $\theta$, or $A_{\mu}$ to give the three
Euler-Lagrange equations.  When $A_{\mu}$ is replace by equation
(\ref{gauge}), the equation involving the derivatives with respect to
$\theta$ is immediately satisfied, which is a consequence of our earlier
choice of $\theta = \phi$.
The covariant derivative $\nabla_{\mu}$ requires
the metric of the space-time to be specified.

We start by assuming that the space-time defined by the self-gravitating
string has general cylindrical symmetry.  The metric has the form:
\be
ds^{2} = -e^{2(K-U)}(dt^{2} - dr^{2}) + e^{-2U}W^{2}d\phi^{2} +
e^{2U}dz^{2}
\ee
where U, K and W are unknown functions of r only and will be solved for in
our calculations.

Using this metric, the Euler-Lagrange equations become:
\beqa
e^{2(U-K)} \left(R'' + R'\frac{W'}{W}\right) - RP^2\frac{e^{2U}}{W^2} -
\frac{dV}{dR} = 0 \label{r}\\
e^{2(U-K)} \left(P'' - P'\left(\frac{W'}{W} - 2U'\right)\right) - e^2R^2P =
0 \label{p}
\eeqa
where $ ' \equiv \partial/\partial r$.

The energy-momentum tensor for the Lagrangian (\ref{lag}) is given by:
\be
T_{\mu\nu} = \frac{2}{\sqrt{-g}}\left\{\frac{\partial (\sqrt{-g}L)}{\partial
g^{\mu\nu}} - (\frac{\partial (\sqrt{-g}L)}{\partial
g^{\mu\nu}_{\;\;,\gamma}})_{,\gamma}\right\}
\ee
The components of $T^{\mu}_{\nu}$ are the following:
\beqa
T^t_t & = & T^z_z = -\frac{e^{2(U-K)}}{2}\left[R'^2 +
\frac{e^{2K}}{W^{2}}R^2P^2
+ \frac{e^{2U}}{W^2e^2}P'^2 + 2Ve^{2K-2U}\right] \\
T^r_r & = & \frac{e^{2(U-K)}}{2}\left[R'^2 - \frac{e^{2K}}{W^{2}}R^2P^2
+ \frac{e^{2U}}{W^2e^2}P'^2 - 2Ve^{2K-2U}\right] \\
T^{\phi}_{\phi} & = & \frac{e^{2(U-K)}}{2}\left[-R'^2 +
\frac{e^{2K}}{W^{2}}R^2P^2
+ \frac{e^{2U}}{W^2e^2}P'^2 - 2Ve^{2K-2U}\right]
\eeqa
The Einstein equations, $G^{\mu}_{\nu} = 8\pi T^{\mu}_{\nu}$,
that couple the string stress-energy to its space-time geometry become:
\beqa
-2\frac{W''}{W} + 2K'\frac{W'}{W} - 2U'^2  = 8 \pi
\left[R'^2 + \frac{e^{2K}}{W^{2}}R^2P^2 + \frac{e^{2U}}{W^2e^2}P'^2
+ 2Ve^{2K-2U}\right] \label{t} \\
2K'\frac{W'}{W} - 2U'^2 = 8 \pi \left[R'^2 - \frac{e^{2K}}{W^{2}}R^2P^2
+ \frac{e^{2U}}{W^2e^2}P'^2 - 2Ve^{2K-2U}\right] \label{rho} \\
2K'' + 2U'^2 = 8 \pi \left[-R'^2 + \frac{e^{2K}}{W^{2}}R^2P^2
+ \frac{e^{2U}}{W^2e^2}P'^2 - 2Ve^{2K-2U}\right] \label{phi} \\
-2\frac{W''}{W} + 4U'\frac{W'}{W} + 4U'' - 2U'^2 - 2K'' = 8 \pi
\left[R'^2 + \frac{e^{2K}}{W^{2}}R^2P^2 + \frac{e^{2U}}{W^2e^2}P'^2
+ 2Ve^{2K-2U}\right] \label{zco}
\eeqa
These equations can be rearranged to give a simpler set of equations.
The first simplification comes from equations (\ref{t})
and (\ref{zco}), from which we obtain $K' = 2U' + \frac{C_1}{W}$.
Since $W(0) = 0$ (see initial conditions), we set $C_1 = 0$ otherwise
$K'(0)\;\rightarrow\;\infty$ and we run into numerical problems (assuming
$U'$ remains finite at the origin).
Therefore, we directly obtain a relationship
between $K$ and $U$ which is of the form $K = 2U + C_2$,
where a new constant of integration is introduced.  Since all solutions
may be transformed such that $K(0) = U(0) = 0$, we set $C_2 = 0$ and
therefore $K = 2U$.

By subtracting equation (\ref{t}) from (\ref{rho}) we obtain an expression for
$W''$:
\be
W'' = -8\pi\left(\frac{e^{2K}R^2P^2}{W} + 2Ve^{2(K-U)}W\right) \label{w}
\ee
Using $K' = 2U'$, and (\ref{rho}) and (\ref{phi}) gives for $U''$:
\be
U'' = 4\pi\left(\frac{P'^2e^{2U}}{e^2W^2} - 2Ve^{2(K - U)}\right) -
U'\frac{W'}{W} \label{u}
\ee
The field equations are integrable only if the conservation equations of energy
and momentum $T^{\mu\nu}_{\;\;  ;\nu} = 0$ are satisfied.  These conservation
laws often give, analogous to the first integrals of classical mechanics,
an important indication of how to solve the field equations.  Using the
energy-momentum tensor given above, the only non-vanishing component of the
conservation equations is:
\beqa
R'\left(e^{2(U-K)}\left(R''+R'\frac{W'}{W}\right) -
RP^2\frac{e^{2U}}{W^2} - \frac{dV}{dR}\right) + \nonumber \\
P'\frac{e^{2U}}{e^2W^2}\left(e^{2(U-K)}\left(P''-P'\left(\frac{W'}{W}
-2U'\right)\right) - e^2R^2P\right) = 0
\eeqa
Note that this equation is a linear combination of
equation (\ref{r}) and (\ref{p}) and therefore indicates that one of
equation (\ref{r}) and (\ref{p}) can be taken as redundant.
Since the conservation of
energy equation was derived from the Einstein equations, one of these
equations can
also be taken as redundant.  We will take equation
(\ref{rho}) as the redundant equation, which we will nevertheless
continue using as a consistency check for our numerical integration, and
keep both equations (\ref{r}) and (\ref{p}).  Equation (\ref{rho}) can be
rewritten as:
\be
R'^2 = \frac{1}{4\pi}\left(K'\frac{W'}{W} - U'^2\right) +
\frac{e^{2K}}{W^2}R^2P^2 - \frac{P'^2e^{2U}}{e^2W^2} + 2Ve^{2(K - U)}
\label{rp}
\ee
A complete set of
equations consists of equations (\ref{r}), (\ref{p}), (\ref{w}), and (\ref{u}),
with equation (\ref{rp}) above.  Similar results have been obtained by
Garfinkle \cite{garfinkle} and Laguna-Castillo and Matzner \cite{matzner}
(hereafter Garfinkle and LCM respectively).

Putting $V = \alpha\frac{e^2}{8}(R^2 - \eta^2)^2$, $K = 2U$,
$e^U = X$, and rescaling $R$ by $\eta$, $W$ and $r$ by $\sqrt{8}/\eta e$
gives us the final set of equations to solve:
\beqa
X'' & = & \frac{1}{2}\pi\eta^2X^3\left(\frac{P'^2}{W^2} - 16\alpha(R^2 -
1)^2\right) + X'\left(\frac{X'}{X} - \frac{W'}{W}\right) \label{z}\\
W'' & = & -8\pi\eta^2X^2\left(X^2\frac{R^2P^2}{W} + 2\alpha W(R^2 - 1)^2\right)
\label{nw}\\
R'' & = & X^4\frac{RP^2}{W^2} + 4\alpha X^2R(R^2 - 1) - R'\frac{W'}{W}
\label{nr}\\
P'' & = & 8X^2R^2P + P'\left(\frac{W'}{W} - 2\frac{X'}{X}\right) \label{np}\\
R'^2 & = & \frac{1}{4\pi\eta^2}\frac{X'}{X}\left(2\frac{W'}{W} -
\frac{X'}{X}\right) + X^2\left(X^2\frac{R^2P^2}{W^2} - \frac{P'^2}{8W^2} +
2\alpha(R^2 - 1)^2\right) \label{nrp}
\eeqa
Notice that the ``$e$''s all disappear from the equations.  It can be
interpreted as a scaling factor.  We are left with a 2-parameter set of
equations instead of 3.  The units that we chose to adopt in this paper
are the natural units: $c = \hbar = G = 1$.  In the original equations,
$R$ had units $[M]$, $W$ and $r$, $[L]$, and $P$ and $X$, $[1]$.
After rescaling, the units of $R$,$W$ and $r$ have become $[1]$
like the other variables.  The dimensions of $\eta$, and $e$, $\lambda$
and $\alpha$ are $[M]$ and $[1]$ respectively.  According to these units,
the energy density will be expressed in terms of $[M]^4$.

\section{Integration Method and Solutions}

For any given set of initial conditions, the differential equations
determine the behavior of $R$, $P$, $X$ and $W$ as a function of $r$.
Although all solutions obtained from any initial set of conditions are
valid solutions of the differential equations,
we will examine in this work only solutions that exhibit the
particular asymptotic behaviour, for which
$\lim_{r\rightarrow\infty} R = 1$ and $\lim_{r\rightarrow\infty} P = 0$.
These conditions are derived from the requirement of finite action.
Such solutions will be denoted  as ``acceptable'' solutions.  They
have the structure of a trapped vortex or string.
An example of an ``acceptable'' solution
and an ``unacceptable'' solution are portrayed in figure 1.

A Taylor-series method was used to numerically integrate the set of five
coupled differential equations.
This method was chosen in part because of the flexibility of implementation
at various orders, which allowed the same computer code to be used for both
rapid exploration of parameter space followed by a more detailed study at
a higher order of integration. In addition the use of a Taylor scheme,
with the production of the higher order Taylor coefficients, allows direct
calculation of many of the geometrical objects of interest (such as the
Christoffel symbols or curvature tensors),
without any need for data fitting or divided difference differentiations.
Because of the complexity of the functions to be integrated, we used the
REDUCE algebraic computing system to produce the mathematical expressions
for the high order derivatives needed for the numerical integration.  The
transformation of the algebraic expressions to build the Taylor integration
program in the C source code was done using the SCOPE package in REDUCE.
The code so generated was verified to be correct by running the C source code
back through REDUCE. This latter check is particularly important when using
automated code production systems.
Several consistency checks were applied to our solutions to ensure
accuracy of the results.
For example, the value of $R'$ computed by the integration scheme was
compared with that obtained from the redundant expression shown in
equation (\ref{nrp}).

The solutions of the differential equations are uniquely determined by
a set of initial conditions.
Recall that the variable $U$ has been defined such that $U(0) = 0$ and
therefore $X(0) = 1$.
We also require that $R(0) = 0$, i.e. the axis is the region of false
vacuum where the potential attains its local maximum.  In order to define an
axis, we must have $W(0) = 0$.
Previous studies have required regularity on the axis and have imposed the
condition that $\lim_{r \rightarrow 0} g_{\phi\phi}/r^2 = 1$
(Garfinkle \cite{garfinkle}).  This means, in terms of our choice of
coordinates, that $W'(0) = 1$.  However, it is
not clear that this condition is justified.
For now, we will set $W'(0) = 1$ and will return to the case $W'(0) \neq 1$
in the next section.

By asking that all derivatives be finite at the origin, we require that
$P(0) = W'(0)$ and $X''(0) =
\frac{\pi\eta^2}{4}\left(\frac{P''(0)^2}{W'(0)^2} - 16\alpha\right)$
while $W'(0)$, $R'(0)$, $P''(0)$ are undertermined and $X'(0)$, $W(0)$,
$W''(0)$, $R''(0)$, $P'(0)$ vanish.
An ``acceptable'' solution is therefore derived from those initial conditions,
given the five free parameters
$\eta$, $\alpha$, $W'(0)$, $R'(0)$, and $P''(0)$.

The ``acceptable'' solutions form a particular subset of all solutions.
The ``goodness'' of a solution is the deviation from the correct
asymptotic structure and is measured by computing $\sqrt{R'^2 + P'^2}$
at large $r$.
In agreement with Shaver, an ``acceptable'' solution is found
at $\alpha = 1.0$, $\eta = 0.19947106$, $W'(0) = 1.0$, $R'(0) = 1.4586085$
and $P''(0) = -4.0$ with a deviation of 0.00003625 measured at
$r_{max} = 4.0$.   From the precision on the parameters, one can see
that to find an ``acceptable'' solution requires a very fine tuning of the
initial parameters.  A small deviation in one of the input parameters
causes a large deviation at $r_{max}$ and gives an ``unacceptable'' solution.
Finding ``acceptable'' solutions is therefore very difficult.  Because of the
``sharpness'' of the valley of deviation of an ``acceptable''
solution (see figure 2 and 3), or, in other words, because of the precision
required for the initial parameters, it is impossible to search randomly
all parameter space in a reasonable time and hope to fall on
an ``acceptable'' solution.
The approach we took for searching for ``acceptable'' solutions
relies on the perturbation method of a known ``acceptable'' solution.  By
manually tuning the initial parameters for a perturbed solution and
by using an extrapolation method to find the next ``acceptable'' solution, we
were able to step away from the known solution and
find the other ``acceptable''
solutions in the parameter space of all solutions.  The subset obtained
using this method covers the surfaces shown in figure 4 and 5.
The ranges of the $\alpha$ and $\eta$ parameters are such that
$\alpha$ covers the region from $0.001$ ($\alpha$ cannot be zero)
to $2.5$ and $\eta$, from $0.0$ to $0.2$.

For each new solution
(from now on we will use ``solution'' to mean ``acceptable'' solution)
we can compute the metric tensor, angular
deficit $\alpha_D$, energy-momentum tensor, Weyl tensor and
Kretschmann scalar using the Taylor coefficients.
We now examine
the change in the physical properties of the string as one moves
in parameter space.  A representative sample of solutions are
shown in figure 6.   The angular deficit $\alpha_D$ (not to be
confused with our input parameter $\alpha$) is defined by
$\Delta\phi = 2\pi (1 - \alpha_D)$,
where $\alpha_D$ is the derivative of $\sqrt{g_{\phi\phi}}$
with respect to $r$.  For our metric,
$\alpha_D = (W' X - W X')/ X^2$.  If $X$ and $X'$
are set equal to 1, we recover Shaver's case.
The first families of solutions, those with $\alpha$ constant, have
$P''(r=0)$ constant as well and show the same energy-momentum
tensor.  The special case where $\alpha = 1.0$ and $P''(0) = -4.0$
yield solutions which have no angular and radial energy-momentum components.
This special case was studied in detail in Shaver.
It was found that the $\alpha_D$ depends on $\eta$ and that
keeping $\alpha_D$ positive required
that $\eta^2 < 4\pi$.  The maximum value for $\eta$ for these
particular solutions is therefore $\sim 0.2821$ and relevant functions for
this solution are displayed in
figure 6.  The angular deficits measured for this set of solutions
are in agreement with Shaver and LCM.   For all computed
energy-momentum tensor components, our values agree with Shaver and
Garfinkle but disagree by a factor of $10$ from the values obtained
by LCM.  It was also noted that LCM obtain a value of $3\%$, which
is also what we obtain, for their value of $1 - e^{A}$ (in our case,
$1 - X^2$) at $\eta = 0.01$ and $\alpha = 1/4$ (corresponding to our
value of $\alpha = 4$) but quote a value of $0.03\%$ in the text.
It should be noted that a factor of $100$ in this expression does
make this solution meaningfully different from Minkowski spacetime.

For the same families of solution, we also notice, in agreement with
LCM, that the angular deficit and Weyl tensor are small for values
of $\eta < 0.01$ (see figure 6) and get larger and more
important for $\eta > 0.01$ with increasing $\eta$.
We also observe that the angular deficit
and Weyl tensor increase with $\alpha$.
%This implies that the maximum
%value of $\eta$ depends on $\alpha$ which means that $\eta_{max}$
%can be pushed to a value as high as $0.2878$ for $\alpha = 0.8$.
Changing the value of the symmetry breaking Higgs field $\eta$ is like
changing the shape of the potential.  By increasing $\eta$, we are allowing
the potential to be wider and the defect to contain more energy.
This shows up in an increase in angular deficit and Weyl effect.
The maximum $\eta$ is reached when the angular deficit becomes $2 \pi$
(or $\alpha_D$ becomes zero).

Examination of the solutions in figure 6, showing non-zero $T^r_r$ and
$T^{\phi}_{\phi}$ components, confirms that the assumptions
made in early works on strings
are not always satisfied.  Simply rescaling the potential of the
scalar field by $\alpha$ changes the shape of the energy-momentum
tensor and the spacetime quite dramatically.  The energy-momentum
tensor with zero angular and radial components gives a potential
with $\alpha = 1$. Moving away from that particular state by
scaling the potential up or down, introduces extra components
of the energy-momentum tensor which we can attempt to explain in the
following way.  When the scalar field acquires more potential
energy and this energy becomes dominant, the particles
in the field will show a higher interaction which can be interpreted
as tension in the string.  The negative components of the energy-momentum
tensor appearing in the corresponding solution correspond to a negative
pressure, i.e. a tension in the $r$ and $\phi$ directions.  Decreasing
the potential energy would mean, by the same argument, that the particles
would be less bound together and this is shown as positive components of
the energy-momentum tensor for the low potential solution.  There seems to
exist a transition region where the $\phi$ component of the energy-momentum
tensor changes sign.  This is probably related to the angular
deficit growing larger as $r$ increases and its action on the dynamics of
the scalar field since the $r$ component of the energy-momentum tensor
does not show the same behavior.

\section{Gravitational lensing from cosmic strings}

The solutions we have derived introduce the possibility of the existence
of cosmic strings in our universe which would curve spacetime around them
and therefore provide some gravitational perturbation to their
surroundings.  The way to study this gravitational perturbation is
to follow the null and timelike geodesics in the string spacetime
and examine their behaviour and consequence for cosmological models.

The gravitational perturbation of cosmic strings on light rays produces
real gravitational lensing and brings a physical (in contrast with
geometrical) prediction for observational tests for cosmic strings.
We first examine the null geodesics in the string spacetime
of each solution.

The string metric written in terms of the new coordinates is:
\be
ds^2 = -X^2(r)(dt^2-dr^2) + \frac{W^2(r)}{X^2(r)}d\phi^2 + X^2(r)dz^2
\ee
and the null geodesic equation can be expressed as:
\be
\frac{-E^2}{X^2(r)} + \frac{X^6(r)}{W^2(r)} L^2 (\frac{dr}{d\phi})^2 +
\frac{X^2(r)}{W^2(r)} L^2 = 0
\ee
The geodesics are described in the plane $z = constant$.
We can rewrite this equation in terms of functions of $r_c$,
the closest approach radius where $dr/d\phi = 0$.  This gives:
\be
\frac{d\phi}{dr} = \pm \frac{1}{\frac{W(r_c)}{X(r_c)^2} h \sqrt{h^2-1}}
\ee
where $h = \frac{W}{W(r_c)}/(\frac{X}{X(r_c)})^2$.
The $\pm$ sign in front of the expression on the right hand side of the
equation can be interpreted as the receding part of
the orbit ($+$ sign) and the approaching part of the orbit ($-$ sign).

Therefore, for the receding part of an orbit starting at $r_c$
and finishing at large $r = r_{max}$,
the angle traversed can be computed using the integral:
\be
\Delta \phi = \int_{r_c}^{r_{max}}
\frac{1}{\frac{W(r_c)}{X(r_c)^2} h \sqrt{h^2-1}}
\ee
For the case where $W(r)/X(r) = \alpha_D r$, this integral can be solve
analytically and, as expected, there is no angular deviation in
the conical spacetime:
\be
\Delta \phi =
\frac{1}{\alpha_D} cos^{-1}(\frac{r_c}{r}) \mid_{r_c}^{r_{max}} \label{D}
\ee
Therefore, when $\sqrt{g_{\phi\phi}}$ is proportional to $r$,
the metric is conformally
flat and one can apply a coordinate transformation and interpret the
difference from Minskowski spacetime by the presence of a missing or
surplus angle.  In the case when the same metric coefficient is a
a nonlinear function of $r$, there is no uniform conformal transformation
which will make the metric flat.
Thus the metric has real curvature and the string is a true gravitating
string.

We use an 8-point Gauss-Legendre integration method to compute the
integrated orbit of a photon in the equatorial plane of the string.
Care is required in the integration near the closest approach, due to the
vanishing denominator, but this can be handled easily by using the limiting
form near closest approach as given in equation (\ref{D}).  The orbits are
plotted for a representative string solution and are shown for different
closest approach radius in figure 7.

In discussing the null orbits, it is important investigate the possible
existence of an event horizon or a photon cylinder (in analogy to the
photon sphere for the Shwarzschild spacetime) associated with our
string solutions.
To determine if the solutions we have found have a photon cylinder,
we need to find the zeroes of $dr/d\phi$. Examination of this function
graphically shows that there is exactly one zero for each ray,
that at closest approach, but no others.
Thus there can be no rays which spiral continuously into the string axis,
and thus there is no photon cylinder. We also find that there is no
event horizon by considering the possibility of nullity of the appropriate
Killing vector.

We are also interested in the Weyl tensor for different gravitating
string solutions to determine whether or not there exist tidal effects
near the string, since these can lead to an understanding of the
distortion, shear, and rotation of geodesics near the string.
All non-vanishing components of the Weyl tensor can be obtained from the
following component:
\beqa
C_{trtr} & = & \frac{2X''XW - 6X'^2W + 4X'W'X - W''X^2}{6W} \\
\eeqa
using the relations:
$ C_{rzrz} = - C_{trtr} $,
$ C_{tztz} = -2 C_{trtr} $,
$ C_{t\phi t\phi} = ( \frac{W}{X^2} )^2 C_{trtr} $,
$ C_{\phi z\phi z} = - C_{t\phi t\phi}$,
and $ C_{r\phi r\phi} = 2 C_{t\phi t\phi} $.
The fact that the Weyl tensor has non-zero components
(see figure 6 for representative functions) combined with the
non-vanishing of the Ricci tensor (as seen through the energy-momentum tensor)
leads to the conclusion that these string solutions will produce
significant gravitational lens effects. These effects are in strong
contrast to the simple ``prism'' optical effect introduced by the
traditional vacuum strings where both the Ricci and Weyl tensors are
identically zero. Thus there will be real distortions and amplification
of distant objects seen along lines of sight passing near these strings.
Of course, in the static situation treated here, there can be no
perturbations to the temperatures of distant sources, such as the cosmic
background radiation.

\section{Regularity at the Origin or is $W'(0) \neq 1$}

In starting to look for solutions, we chose to set $W'(0) = 1$.
This has been consistently assumed throughout the literature on
cosmic strings, on the basis that $W'(0) = 1$ implies that the
axis of the string is regular.
On the other hand, there seem to be few arguments that lead to a
justification of this particular assumption of regularity.
We will demonstrate in this section that this assumption of
regularity at the axis of the cosmic string is not necessary
and that in fact, the physics of the spacetime is the same if the
angular deficit is put concentrated near the axis or if it is put at
large $r$.

The first thing we want to do is ensure that there is no real singularity
at the string axis.  By singularity we mean the divergence of the
Kretschmann scalar defined by
$\cal{K}= R^{\sigma \tau \mu \nu}R_{\sigma \tau \mu \nu}$
where $R_{\sigma \tau \mu \nu}$ is the Riemann tensor.
For the general metric given above, the Kretschmann scalar is given by:
\beqa
\cal{K} & = & \frac{4}{X^8W^2}( 3X''^2X^2W^2 - 10X''X'^2XW^2 + 6X''X'W'X^2W \\
& &  - 2X''W''X^3W + 14X'^4W^2 - 22X'^3W'XW + 6X'^2W''X^2W \\
& &  + 11X'^2W'^2X^2 - 6X'W''W'X^3 + W''^2X^4 )
\eeqa
We observe that the Kretschmann scalar and its derivative
do not diverge for any
solutions implying that there is no real singularity at the origin of
the gravitating string.

The solutions we obtain when varying the parameter $W'(0)$
in addition to the other parameters mentioned above have the
same $R'(0)$ (same parameter space as shown in figure 4).
The difference lies in the
value of $P''(0)$, which is simply the value of $P''(0,W'(0)=1)$
multiplied by $W'(0)$.
The solution surface for $P''(0)$ therefore is the same as
figure 5 scaled by $W'(0)$.  This completes the solution
space.  We can now compare solutions for which the only parameter
varied is $W'(0)$.  We find that they exhibit the same
physical bending and therefore are equivalent (see figure 8 and 9).
The angular deficit $\alpha_D$, the $\phi$ component of the
energy momentum tensor and the $\phi$ components of the Weyl tensor
are observed to increase
with $W'(0)$.  This increase is by a factor
$\alpha_D^2$, which is consistent with a simple coordinate
transformation.  This demonstrates that
the physics of the string spacetime
does not depend on the choice of $W'(0)$ and therefore that
the regularity at the axis is not a necessary condition.

\section{Conclusion}
We have shown that the Einstein-Euler-Lagrange equations describing a
gravitating cosmic string can be solved simultaneously and accurately
using a Taylor series method.  This method allows us to also study the
physical properties of the string solutions for any set of initial
parameters we wish to examine.  The question of regularity at the axis of
the string was resolved in this work using physical arguments.
The principal outcome of the work described in this paper is that
there does occur significant gravitational lensing for many of
the solutions found.  These strings may have important consequences in
cosmology since they now are gravitating and are true gravitational lenses.

\section{Acknowledgments}
We would like to thank Eric Shaver and Glenn D. Starkman for many
helpful discussions.
This work has been supported by the Natural Sciences and Engineering
Research Council of Canada through a Postgraduate Scholarship (F.~R.~M.)
and an operating grant (C.~C.~D.).

\begin{figure}[p]
\caption{ The magnitude of the scalar and gauge field
as a function of $r$ and the difference between an ``acceptable'' and
``unacceptable'' solutions. }
\end{figure}

\begin{figure}[p]
\caption{ Contour plot of the measured deviation from the correct
asymptotic structure
around the ``acceptable'' solution with parameters $\alpha = 1.0$,
$\eta = 0.19947106$, $R'(0) = 1.4586085$ and $P''(0) = -4$.  The contours
are plotted for values of the deviation of $0$, $0.5$, $1$, $2$,
$5$, $10$, $20$ and $20000$.  There is a deep valley very near the
``acceptable'' solution were the deviations get smaller than in the
surrounding regions. }
\end{figure}

\begin{figure}[p]
\caption{ Circular scans of the measured deviation shown in figure 2.
Scans are drawn for circles of different radii around
the ``acceptable'' solution.  A complete circle is shown by an angle
going from $0$ to $2 \pi$ as shown on the abscissa.  The narrow valley of
deviations is more apparent in this figure. }
\end{figure}

\begin{figure}[p]
\caption{ Surface representing the subset of ``acceptable'' solution.
The surface covers a range in $\alpha$ of $0.001$ to $2.5$ and a range
in $\eta$ of $0.0$ to $0.2$.  This plot shows the value of $R'(0)$
for a specific choice of $\alpha$ and $\eta$ parameters that will give
an ``acceptable'' solution. }
\end{figure}

\begin{figure}[p]
\caption{ For the same range as given in figure 4, this plot gives the
value of $P''(0)$ for a specific choice of $\alpha$ and $\eta$ parameters
that will give an ``acceptable'' solution. }
\end{figure}

\begin{figure}[p]
\caption{ Shown here are different ``acceptable'' solutions and some of
their physical characteristics.
The first row shows the solution $\alpha = 1.0$,
$\eta = 0.19947106$, $W'(0) = 1.0$, $R'(0) = 1.4586085$ and $P''(0) = -4$,
the second row $\alpha = 1.0$, $\eta = 0.2820947$, $W'(0) = 1.0$,
$R'(0) = 1.213061694$ and $P''(0) = -4.0$, the third row
$\alpha = 0.319$, $\eta = 0.15$, $W'(0) = 1.0$,
$R'(0) = 1.081678$ and $P''(0) = -2.848065$
and the last row $\alpha = $1.5, $\eta = 0.01$, $W'(0) = 1.0$,
$R'(0) = 1.9746463$ and $P''(0) = -4.5213392$. }
\end{figure}

\begin{figure}[p]
\caption{ The orbit of a photon in the equatorial plane of a cosmic string,
plotted in terms of the cylindrical coordinates $r$ and $\phi$, for selected
values of the closest approach radius of the orbit.  The orbits depicted here
are for the solution with parameters $\alpha = 2.0$, $\eta = 0.15$,
$W'(0) = 1.0$, $R'(0) = 1.9955592$ and $P''(0) = -4.923572$. }
\end{figure}

\begin{figure}[p]
\caption{ The orbit of a photon are shown here for the solution with
$\alpha = 1.0$, $\eta = 0.19947106$, $W'(0) = 1.0$,
$R'(0) = 1.4586085$ and $P''(0) = -4.0$. }
\end{figure}

\begin{figure}[p]
\caption{ The orbit of a photon are shown for the solution with
$W'(0) \neq 1.0$, that is with
$\alpha = 1.0$, $\eta = 0.19947106$, $W'(0) = 0.5$,
$R'(0) = 1.4586085$ and $P''(0) = -2.0$.  There is no difference
between this figure and the previous one with $W'(0) = 1.0$. }
\end{figure}


\begin{thebibliography}{99}
\bibitem{shafi}  Q.~Shafi and A.~Vilenkin, {\it Phys.~Rev.~D} {\bf 29},
1870 (1984).
\bibitem{lazarides}  T.~W.~B.~Kibble, G.~Lazarides and Q.~Shafi,
{\it Physics Letters} {\bf 113B}, 237 (1982).
\bibitem{vilenkin}  A.~Vilenkin, {\it Phys.~Rev.~D} {\bf 23}, 852 (1981).
\bibitem{zeldovich}  Ya.~B.~Zel'dovich {\it et al.}, {\it JETP} {\bf 67},
3 (1974).
\bibitem{kibble}  T.~W.~B.~Kibble, {\it J.~Phys.~A} {\bf 9}, 1387 (1976).
\bibitem{gott}  J.~R.~Gott, {\it Astrophys.~J.} {\bf 288}, 422 (1985).
\bibitem{hiscock}  W.~A.~Hiscock, {\it Phys.~Rev.~D} {\bf 31}, 3288 (1985).
\bibitem{raychaudhuri}  A.~K.~Raychaudhuri, {\it Phys.~Rev.~D} {\bf 41},
3041 (1990).
\bibitem{matzner}  P.~Laguna-Castillo and R.~A.~Matzner,
{\it Phys.~Rev.~D} {\bf 36}, 3663 (1987).
\bibitem{laguna}  D.~Garfinkle and P.~Laguna, {\it Phys.~Rev.~D} {\bf 39},
1552 (1989).
\bibitem{shaver}  Eric Shaver, {\it Gen.~Rel.~Grav.} {\bf 24}, 187 (1992).
\bibitem{nielsen}  H.~B.~Nielsen and P.~Olesen, {\it Nucl.~Phys.} {\bf B61},
45 (1973).
\bibitem{garfinkle}  D.~Garfinkle, {\it Phys.~Rev.~D} {\bf 32}, 1323 (1985).
\bibitem{jaffe}  A.~Jaffe and C.~Taubes, {\it Vortices and Monopoles}
(Birkhauser, Boston, 1980).
\end{thebibliography}
\end{document}